\documentstyle[twoside,fleqn,espcrc2]{article}

\newcommand{\AmS}{{\protect\the\textfont2
  A\kern-.1667em\lower.5ex\hbox{M}\kern-.125emS}}

\title{ Electromagnetic Structure of Light Baryons 
 in Lattice QCD }
\author{A.~Duncan\address{Dept. of Physics and Astronomy,University of Pittsburgh,
 Pittsburgh, PA 15260},%
  E.~Eichten\address{Fermilab, P.O. Box 500, Batavia, IL 60510}%
  and 
  H.~Thacker\address{Dept. of Physics, University of Virginia, 
 Charlottesville, VA 22901}}

\begin{document}
\begin{abstract}
A method in which electromagnetic properties of hadrons  are
 studied by direct simulation of dynamical photon effects
 is applied to the extraction of  the isomultiplet structure
of the octet baryons.
Using 187 configurations at $\beta=5.7$ with Wilson action, and
up and down quark masses determined from the meson spectrum,
the nucleon  splitting is found to be $1.55(\pm 0.56\; \rm stat)$ MeV;
the hyperon splittings are found to be $\Sigma^{0}-\Sigma^{+}=2.47\pm
0.39$, $\Sigma^{-}-\Sigma^{0}=4.63\pm 0.36$, $\Xi^{-}-\Xi^{0}=5.68\pm
0.24$ MeV.
Estimated systematic corrections arising from finite volume and
the quenched approximation are included in these results.
\end{abstract}

\maketitle
\section{Introduction}
The fact that the intrinsic up-down quark mass difference is comparable to
typical hadronic electromagnetic energy shifts has made it difficult to 
reliably calculate the octet baryon isomultiplet splittings in the absence
of a systematic treatment of virtual photon effects combined with nonperturbative
QCD contributions. The problem is particularly acute in the case of the proton-neutron
mass difference, where tadpole (quark mass) effects almost completely cancel
the photon cloud contributions (see \cite{Zee} for an early review exhibiting the
massive confusion prevailing in pre-QCD days). 
In this talk, we extend a method recently
used \cite{Estitalk} to extract  electromagnetic contributions to pseudoscalar masses
to the  octet baryon spectrum. The basic idea is to propagate the
quarks through a $U(3) = SU(3)\times U(1)$ field including both dynamical gluon and
 photon effects.The calculation of baryon isomultiplet splittings can then be viewed as a two-step
process:\\
(a) First, the bare quark masses are determined from an analysis \cite{Estitalk} of the
pseudoscalar meson spectrum (including electromagnetic contributions);\\
(b) Secondly, the octet baryon spectrum is computed (again including dynamical
photon effects) and extrapolated to physical values of quark mass (as determined
 in step (a)) and electric charge.

\section{Extraction and Fitting of Baryon Spectrum}

The strategy of the calculation is as follows: quark 
propagators are generated in the
presence of Coulomb gauge background SU(3)$\times$U(1) fields.  
187 gauge configurations, separated by 1000 Monte Carlo sweeps,
 were generated at $\beta=5.7$ on a
$12^3\times 24$ lattice.
Quark propagators are calculated for 4 electric charges
and  3 light quark mass values, and with either a local or smeared source
 (see \cite{Estitalk} for details).
From the resulting 12 quark propagators, 936 independent octet baryon
three-quark combinations can be formed.

 In quenched QCD it is known \cite{sharpe} that
baryon masses are described by a function
of the bare quark masses involving nonanalytic $m_{q}^{3/2}$ 
(as well as linear) terms, and terms involving logarithms
of the quark mass arising from the same hairpin diagrams
familiar in the quenched meson spectrum \cite{bern,sharpmes}.
The latter terms now appear to be extremely small numerically
\cite{hank,volume} : we neglect them throughout.  However,
 we do include  terms of order ${\rm (quark mass)}^{3/2}$.
Thus a general octet baryon mass is written 
\begin{eqnarray}
\label{eq:ChPT}
m_{B} &=& A(e_{q1}, e_{q2}, e_{q3}) + \sum_{i} m_{qi}B_i(e_{q1}, e_{q2}, e_{q3})
 \nonumber \\ 
&~&+ \sum_{i,j}(m_{qi}+m_{qj})^{3/2}C_{ij}(e_{q1}, e_{q2}, e_{q3}) 
\end{eqnarray}
where $e_{q1}, e_{q2}, e_{q3}$ are the three quark charges, and 
$m_{q1}, m_{q2}, m_{q3}$
are the three bare quark masses, defined in terms of the Wilson hopping parameter by
$(\kappa^{-1} - \kappa^{-1}_c)/2a$. (Here $a$ is the lattice spacing.)
Each of the coefficients $A,B_i,C_{ij}$ in (\ref{eq:ChPT}) is then expanded
in powers of the quark charges $e_{q1}, e_{q2}, e_{q3}$, with terms up to fourth
 order for $A$, second order for $B_i$, and with no charge dependence 
 assumed for the nonanalytic $C_{ij}$ terms.
  (\ref{eq:ChPT}) turns out to 
have 30 parameters once all symmetries are exploited.

  We have varied  the baryon mass window (for each
 choice of Euclidean time window used to extract a mass from 
 smeared-local correlators) until the 
 $\chi^2/{\rm dof}$ of the fit to (\ref{eq:ChPT}) was minimized. For example, using a Euclidean time
 window from $t=$5 to $t=$8, the mass window (lattice units) from 
 1.20 to 1.26 was found to contain 74  baryons. Determining the 30
 parameters in (\ref{eq:ChPT}) by fitting this set of masses gave
 a $\chi^2/{\rm dof}$ of 1.33 . By contrast, using the mass window
 from 1.15 to 1.20 (122 baryons), the chi-square fit minimizes at
 $\chi^2/{\rm dof}$=2.16. For each choice of Euclidean time window, we
 have performed the fit to (\ref{eq:ChPT}) using a baryon mass window
 which optimizes the $\chi^2/{\rm dof}$ - the overall optimal fit was found
 for the window $t=$ 5 to 8. One then  determines the
 mass of any given octet baryon by extrapolating to physical values
 of quark mass and charge. The propagators for different electric charge are highly correlated,
 so it is not surprising  that the statistical error on the
 center of gravity of  isomultiplets is considerably larger
 than the error on multiplet splittings.  
\begin{table}
\begin{center}
\caption{Raw lattice results (MeV) for baryon octet ($\beta$=5.7, 12$^3$x24, 187 configurations)}
\begin{tabular}{|c|c|}
\hline
\multicolumn{1}{|c|}{Baryon State}
&\multicolumn{1}{c|}{Window 5-8 ($\chi^2/{\rm dof}$=1.33)}   \\ \hline
\multicolumn{1}{|c|}{Parameters}
&\multicolumn{1}{c|}{$m_{u,d,s},a^{-1}$=3.57,7.10,155,1370} \\ \hline
N  &  935.92$\pm$42.4 \\ 
P  &  933.07 $\pm$42.9  \\ 
N-P & 2.83 $\pm$ 0.56  \\  
$\Sigma^{+}$  & 1171.6 $\pm$ 25.6  \\
$\Sigma^{0}$  & 1175.1 $\pm$ 25.3  \\
$\Sigma^{-}$  & 1179.1 $\pm$ 25.0  \\
$\Sigma^{0}-\Sigma^{+}$  & 3.43 $\pm$ 0.39  \\
$\Sigma^{-}-\Sigma^{0}$  & 4.04 $\pm$ 0.36  \\
$\Sigma^{+}+\Sigma^{-}-2\Sigma^{0}$  & 0.61 $\pm$ 0.19 \\
$\Xi^{-}$  &  1312.9 $\pm$ 14.5 \\
$\Xi^{0}$  &  1308.2 $\pm$ 14.6 \\
$\Xi^{-}-\Xi^{0}$  &  4.72 $\pm$ 0.24 \\
$\Lambda^{0}$  & 1098 $\pm$ 52  \\
\hline
\end{tabular}
\end{center}
\end{table}
 Note, in connection with the raw lattice results quoted in Table 1, the following :\\
(1)  The quark mass parameters and lattice
scale assumed in generating  masses for each of the fitting window choices
are shown in the second row. The up and down quark masses are those obtained
from the pseudoscalar spectrum (these will depend on the scale). The strange quark mass is known to fall at a higher value when determined from
the baryon spectrum (due to
 discretization and quenched errors), so we have chosen to fix it using the
center of gravity of the $\Xi$ hyperon, which has the smallest statistical errors
in our analysis. The center of gravity of the $\Sigma$ multiplet and the $\Lambda$
mass are then predictions of the analysis. \\
(2) The lattice scale has been fixed in each case by requiring the nucleon center
of gravity to sit at (roughly) the physical value. \\

\begin{table*}[hbt]
\caption{Final results for baryon octet splittings ($\beta$=5.7, 12$^3$x24, 187 configurations)}
\begin{tabular}{|c|c|c|c|c|c|}
\hline
\multicolumn{1}{|c|}{Level Splitting}
&\multicolumn{1}{c|}{Raw Lattice}
&\multicolumn{1}{c|}{Finite Volume}
&\multicolumn{1}{c|}{Meson Cloud}
&\multicolumn{1}{c|}{Total Lattice}
&\multicolumn{1}{c|}{Physical}   \\ \hline
N - P & 2.83 $\pm$ 0.56&  -0.75  &  -0.53  & 1.55 $\pm$ 0.56 &   1.293 \\
$\Sigma^{0}-\Sigma^{+}$  & 3.43 $\pm$ 0.39 & -0.80 &  -0.16 & 2.47 $\pm$ 0.39 &  3.18 $\pm$ 0.1 \\
$\Sigma^{-}-\Sigma^{0}$  & 4.04 $\pm$ 0.36  & +0.86  &  -0.27 & 4.63 $\pm$ 0.36  & 4.88 $\pm$ 0.1\\
$\Sigma^{+}+\Sigma^{-}-2\Sigma^{0}$& 0.61 $\pm$ 0.19& +1.66 &  -0.11 & 2.16 $\pm$ 0.19 & 1.70 $\pm$ 0.15 \\
$\Xi^{-}-\Xi^{0}$  &  4.72 $\pm$ 0.24 & +0.86  &  +0.10  & 5.68 $\pm$ 0.24 & 6.4 $\pm$ 0.6  \\
\hline
\end{tabular}
\end{table*}

\section{Finite Volume and Quenched Corrections}

 With  massless physical degrees of freedom we expect finite 
 volume effects which fall  as inverse powers of the lattice size.  These
corrections can be studied directly on the lattice by repeating the calculations
on lattices of varying physical volume. Here we estimate them by
using the known dominance  of the Born contribution to the
dispersive evaluation of the Cottingham  formula. 
 Single photon exchange can be written as a sum of an electric
and magnetic contribution to hadronic self-energies- the electric term takes the form
\begin{eqnarray}
\label{eq:finvol1}
 \delta m_{\rm el} &=& 2\pi\alpha m \frac{1}{L^3}\sum_{\vec{q}\neq 0}
 \frac{G_{E}(q)^2}{|q|} \{\frac{2}{q^2 +4m^2}~~   \nonumber  \\  
 &~&~~~~~~~~+\frac{1}{2m^2}(\sqrt{1+\frac{4m^2}{q^2}}-1)\}  
\end{eqnarray}
where the momentum vectors $\vec{q}$ are the  discretized bosonic photon
momenta for the finite LxLxL lattice. Using (\ref{eq:finvol1}) one can estimate the
finite volume corrections to baryon masses on our L=12 lattice- they are indicated 
in column 3 of Table 2, together with our final estimate (including the
finite volume correction as well as quenched error estimate- see below)
for the  baryon mass in column 5.

 Processes in which mesons are emitted and reabsorbed from a baryon
 include graphs with internal quark loops and are known \cite{leutrev} to result in a small but nonnegligible shift in isospin splittings. For example, 
in the static limit where the nucleon mass is infinite,  the pion cloud {\em decreases}
the neutron-proton splitting by an amount (in the infinite volume limit)
 0.43$\Delta M_0$, where $\Delta M_0$ is the nucleon splitting in the
 absence of a virtual pion cloud (a fully relativistic evaluation gives
 0.41$\Delta M_0$). 
   We shall use a static approximation  but include the effects of all octet pseudoscalar mesons (assuming SU(3)
 symmetry with a $d:(f+d)$ ratio of 0.62). Discretizing the second order
 shift formula  (see \cite{leutrev}) on a LxLxL lattice, one may estimate the 
 meson cloud shift for the particular lattices used. 
 Since the meson cloud shift includes contributions from quenched 
 nonplanar graphs in the
 cases where the emitted meson only contains valence quarks of the external
 baryon,  these estimates  are only a rough indication of the
  magnitude and sign  (probably, an overestimate),
  of the quenched correction. Setting L=12 and
using a lattice scale $a^{-1}$=1370 MeV, together with the quenched masses
from column 2 of Table 1, we obtain the meson cloud shifts given in column 4
 of Table 2. The lattice results, corrected for finite volume 
 and meson cloud effects, are given in column 5,
 and the physical values in column 6.

 The results in Table 2 (which must still  be corrected for finite lattice
spacing effects) suggest that this first  evaluation, on a fairly
 coarse lattice, already reproduces - almost quantitatively- the isomultiplet
 pattern  of the octet baryons.
 Of course, the extremely delicate level of baryon fine structure being considered here
 requires a detailed study
 of all systematic effects, with improved statistics on larger lattices. An upcoming
 run will work with lattices of varying physical volume, using  improved
 (to O(a)) action to minimize lattice discretization errors- an important check given the known
 strong a-dependence in off-shell defined continuum quark mass parameters \cite{pbm}.

We thank  George Hockney
 for continuing contributions to our effort. 



\end{document}